\documentclass[11pt,a4paper]{article}
\pdfoutput=1 
\usepackage{jcappub}

\usepackage[normalem]{ulem}
\usepackage{comment}
\usepackage{ifpdf}
\usepackage{amsfonts}
\usepackage{mathtools}
\usepackage{braket} 
\usepackage{tensor} 
\usepackage{bm} 
\usepackage{slashed}
\usepackage{enumerate}
\usepackage{feynmp}
\usepackage{multirow}
\usepackage{subfigure}
\usepackage{dcolumn}
\usepackage{xcolor}

\begin{document}


\title{Primordial Origin of Supermassive Black Holes from Axion Bubbles}
\author[a]{Kentaro Kasai}
\author[a,b]{, Masahiro Kawasaki}
\author[c,d]{, Naoya Kitajima}
\author[d]{, \\Kai Murai}
\author[a]{, Shunsuke Neda}
\author[d]{, Fuminobu Takahashi}
\affiliation[a]{ICRR, University of Tokyo, Kashiwa, 277-8582, Japan}
\affiliation[b]{Kavli IPMU (WPI), UTIAS, University of Tokyo, Kashiwa, 277-8583, Japan}
\affiliation[c]{Frontier Research Institute for Interdisciplinary Sciences, Tohoku University, Sendai, 980-8578 Japan}
\affiliation[d]{Department of Physics, Tohoku University, Sendai, 980-8578 Japan}

\abstract{%
We study a modification of the primordial black hole (PBH) formation model from axion bubbles.
We assume that the Peccei-Quinn scalar rolls down in the radial direction from a large field value to the potential minimum during inflation, which suppresses the axion fluctuations and weakens the clustering of PBHs on large scales.
We find that the modified model can produce a sufficient number of PBHs that seed the supermassive black holes while avoiding the observational constraints from isocurvature perturbations and angular correlation of the high-redshift quasars.
}


\emailAdd{kkasai@icrr.u-tokyo.ac.jp}
\emailAdd{kawasaki@icrr.u-tokyo.ac.jp}
\emailAdd{naoya.kitajima.c2@tohoku.ac.jp}
\emailAdd{kai.murai.e2@tohoku.ac.jp}
\emailAdd{neda@icrr.u-tokyo.ac.jp}
\emailAdd{fumi@tohoku.ac.jp}

\begin{flushright}
    TU-1213
\end{flushright}

\maketitle

\section{Introduction}
\label{sec: intro}
Recently, the pulsar timing array experiments, NANOGrav~\cite{NANOGrav:2023hfp}, EPTA~\cite{EPTA:2023fyk} with InPTA data, PPTA~\cite{Reardon:2023gzh}, and CPTA~\cite{Xu:2023wog}, suggested the existence of a low-frequency gravitational-wave background.
One of the most promising candidates for its source is inspiraling supermassive black holes (SMBHs), and hence the scenario of SMBH formation and its distribution are attracting more attention in astrophysics, cosmology, and particle physics.
The analysis of the James Webb Space Telescope data has identified active galaxy nuclei at high redshifts up to $z = 10.6$ (e.g., Refs.~\cite{Onoue:2022goe,Kocevski:2023,Barro:2023,Ubler:2023,CEERSTeam:2023qgy,Harikane:2023,Maiolino:2023zdu,Maiolino:2023bpi,Goulding:2023gqa}), which suggests the existence of SMBHs at high redshifts.
One may expect that such early SMBHs are seeded by Pop~III stars.
Pop III stars are formed from primordial gas and their mass is typically $\sim 10\,\text{--}\,10^3M_\odot$ according to the hydrodynamical simulation~\cite{Hirano:2013lba}.
Most part of them finally become black holes at the end of their evolution.
However, if SMBHs are formed from Pop III stars, the observed SMBHs at high redshift require more efficient accretion than the Eddington limit.
An alternative promising candidate for the seeds of SMBHs is primordial black holes (PBHs).
In contrast to astrophysical black holes, PBHs can explain the wide range of masses including $\gtrsim 10^4 M_\odot$.
In the standard PBH formation scenarios, PBHs are formed from large density perturbations generated by inflation.
However, the $\mu$-distortion of the CMB spectrum stringently constrains the formation of PBHs with masses larger than $\sim 10^4 M_\odot$ from (nearly) Gaussian density perturbations~\cite{Chluba:2012we,Kohri:2014lza}.
This compels us to consider PBH formation models where PBHs are produced from highly non-Gaussian density fluctuations.
In this paper, we study the PBH formation from axion bubbles~\cite{Kitajima:2020kig} focusing on the PBHs that seed SMBHs.
This scenario is analogous to that using the Affleck-Dine mechanism~\cite{Dolgov:2008wu,Blinnikov:2016bxu,Hasegawa:2017jtk,Hasegawa:2018yuy,Kawasaki:2019iis,Kasai:2022vhq} and predicts PBH masses depending on the axion decay constant $f_a$.
For $f_a=10^{16}\,\mathrm{GeV}$, PBHs with masses of $\mathcal{O}(10^4) M_\odot$ are formed, which can be the seeds of SMBHs.
Recently, we showed that this model leads to strong PBH clustering and large isocurvature perturbations are produced, from which we obtained a stringent constraint on the model~\cite{Kasai:2023ofh}.
Following the procedure developed in Ref.~\cite{Kawasaki:2021zir}, we found that the clustering of PBHs and their byproduct, axion miniclusters (AMCs), is very strong, and the PBH fraction in dark matter is constrained as $f_\mathrm{PBH}\lesssim 7 \times 10^{-10}$
for $f_a=10^{16}\,\mathrm{GeV}$.
Since the observational amount of SMBHs is estimated as $f_\mathrm{SMBH}\sim 3\times 10^{-9}$ for $M_\mathrm{SMBH} > 10^6 M_\odot$~\cite{Willott:2010yu}, this scenario requires a significant matter accretion into PBHs to explain the origin of the SMBHs.
Moreover, there is another observational constraint from the angular correlation of quasars.
Refs.~\cite{Shinohara:2021psq,Shinohara:2023wjd} show that the strong clustering of PBHs leads to a large angular correlation of the observed quasars at $z\sim 6$, which rules out the PBH scenario with strong clustering.
Therefore, we have to alleviate the clustering of PBHs for the scenario to explain SMBHs successfully.
In this paper, we modify the model proposed in Ref.~\cite{Kitajima:2020kig} so that it can evade the strong observational constraints above.
Those constraints mainly come from axion fluctuations on large scales ($\gtrsim 1$\,Mpc).
Thus, we can avoid the constraints if the large-scale axion fluctuations are suppressed.
In this paper, we consider the Peccei-Quinn (PQ) scalar that slowly rolls down from a large field value to the potential minimum during inflation.
Since the axion fluctuations are inversely proportional to the radial amplitude of the PQ scalar, the spectrum of the axion fluctuations is blue-tiled and hence suppressed on large scales~\cite{Kasuya:2009up,Chung:2015pga}.
Finally, we find that a sufficient amount of seed PBHs for the SMBHs can be produced while avoiding the observational constraints including that from the angular correlation.
The rest of this paper is organized as follows.
In Sec.~\ref{sec: SMBHm}, we review the scenario of the axion bubbles proposed in Ref.~\cite{Kitajima:2020kig} and evaluate the abundance and clustering of PBHs in our modified model.
In Sec.~\ref{sec: themo}, we apply the observational constraints to our model and discuss the viability of the model.
Finally, Sec.~\ref{sec: concl} is devoted to the conclusion.
\section{PBH formation from axion bubbles}
\label{sec: SMBHm}
\subsection{Original model}
\label{subsec: Revie}
First, we review the original model of PBH formation from QCD axion bubbles~\cite{Kitajima:2020kig}.
We assume that the PQ symmetry is spontaneously broken before or during inflation.
The QCD axion $\phi$ is a massless spectator field, and it acquires fluctuations during inflation.
After inflation, the axion obtains a mass through non-perturbative effects of QCD.
The QCD axion couples to gluons through
\begin{align}
    \mathcal{L}
    \supset
    \frac{g^2}{64\pi^2}\frac{\phi}{f_a}\epsilon_{\mu\nu\rho\sigma}G^{a\mu\nu}G^{a\rho\sigma}
    ,
\end{align}
where $\epsilon_{\mu\nu\rho\sigma}$ is the totally anti-symmetric tensor in four dimensions, $G^{a\mu\nu}$ is the field strength tensor of gluons, and $g$ is the strong gauge coupling.
The strength of this coupling is characterized by the decay constant $f_a$.
Due to non-perturbative QCD effects, this interaction induces the effective potential for the axion given by\footnote{The precise form of the potential in the low energy deviates from the simple cosine function, but this difference is not relevant for the following discussion.}
\begin{align}
    V_\mathrm{QCD}(\phi)
    =
    m_a^2(T)f_a^2\left(1-\cos{\frac{\phi}{f_a}}\right).
\end{align}
According to the lattice QCD calculation, the axion mass depends on the temperature $T$ as
\begin{align} 
    m_a(T)\simeq
    \begin{cases}
        0.57\,{\rm{neV}}
        \left(\frac{10^{16}\,\rm{GeV}}{f_a}\right)
        \left(\frac{0.15\,{\rm GeV}}{T}\right)^{\frac{c}{2}}  
        &
        T>0.15\,{\rm{GeV}}  
    \\[0.5em]
        0.57\,{\rm{neV}}\left(\frac{10^{16}\,{\rm GeV}}{f_a}\right)       
        &
        T<0.15\,{\rm{GeV}}
    \end{cases}
    \ ,
    \label{eq: axion mass}
\end{align}
where $c = 8.16$~\cite{Borsanyi:2016ksw}. 
Note that we set $\phi = 0$ as the strong CP conserving point without loss of generality.
For our purpose, we need an additional PQ breaking term that becomes relevant before the QCD phase transition. The main idea of the QCD axion bubbles does not require a specific UV completion for such an additional PQ breaking term, but for concreteness we consider the Witten effect~\cite{Witten:1979ey,Fischler:1983sc}. 
For this purpose, we consider that $\phi$ is coupled with a hidden $\mathrm{U}(1)_H$ gauge field through
\begin{align}
    \mathcal{L} 
    \supset 
    - \frac{\alpha_H}{16 \pi} \left(
        N_H \frac{\phi}{f_a} + \theta_H
    \right)
    \epsilon_{\mu \nu \rho \sigma}
    F_H^{\mu \nu} F_H^{\rho \sigma}
    \ ,
    \label{eq:axion_hiddenU1_coupling}
\end{align}
where $\theta_H$ is the $\theta$-parameter of the $\mathrm{U}(1)_H$ gauge symmetry, and $F_H^{\mu\nu}$ is the field strength of the $\mathrm{U}(1)_H$ gauge field.
This coupling is characterized by the constant $\alpha_H$ and the domain wall number $N_H$. We take $N_H=2$ in this paper.
The non-zero $\theta$ parameter induces an electric charge on hidden magnetic monopoles, which is called the Witten effect~\cite{Witten:1979ey,Fischler:1983sc}.
Since such monopoles are accompanied with the axion-dependent electric field, the background monopoles induce a potential for the axion.
This potential results in the axion mass given by
\begin{align}
    m_{a, W}^2=\frac{\alpha_H N_H}{16\pi^2}\frac{n_M}{r_cf_a^2}
    \ ,
    \label{eq: induced mass}
\end{align}
where $n_M$ is the total number density of monopoles and anti-monopoles, and $r_c$ is the radius of the monopole core.
The potential induced by the Witten effect has the $2\pi f_a/N_H$-periodicity of $\phi$.
Now, we briefly summarize the original scenario. 
\begin{enumerate}
    \item During inflation, the axion field value spatially fluctuates due to quantum fluctuations.
    As a result, the axion field at a point $x$ follows a Gaussian probability distribution $P(N, \phi)$, where $N$ is the e-folding number.
    \item After inflation, $U(1)_H$ monopoles are produced, which induces the axion potential through the Witten effect~\cite{Kawasaki:2015lpf,Nomura:2015xil}.
    Then, the axion field value settles down to one of the minima of the potential,\footnote{%
    Here we assume that, depending on the value of the axion, only monopoles in one branch are produced. The actual fraction of monopoles produced in different branches should be determined by dedicated numerical calculations.
    }
    \begin{align}
       \phi_\mathrm{min}^{(n)}=(-\theta_H+2\pi n)\frac{f_a}{2}
       ~~~~~(n=0,1),
       \label{eq: minima}
    \end{align}
    depending on the field value, which follows $P(N, \phi)$ at the end of inflation.
    \item After the axion field settles down to the minima, the $U(1)_H$ symmetry is spontaneously broken to produce cosmic strings connecting monopoles and anti-monopoles. 
    The monopoles and anti-monopoles then annihilate due to the tension of cosmic strings, and the induced potential vanishes.
    \item Around the QCD scale, the axion starts to oscillate due to the effective potential induced by the QCD instanton effect.
\end{enumerate}

\begin{figure}[ht]
    \centering
    \includegraphics[width=.4\textwidth ]{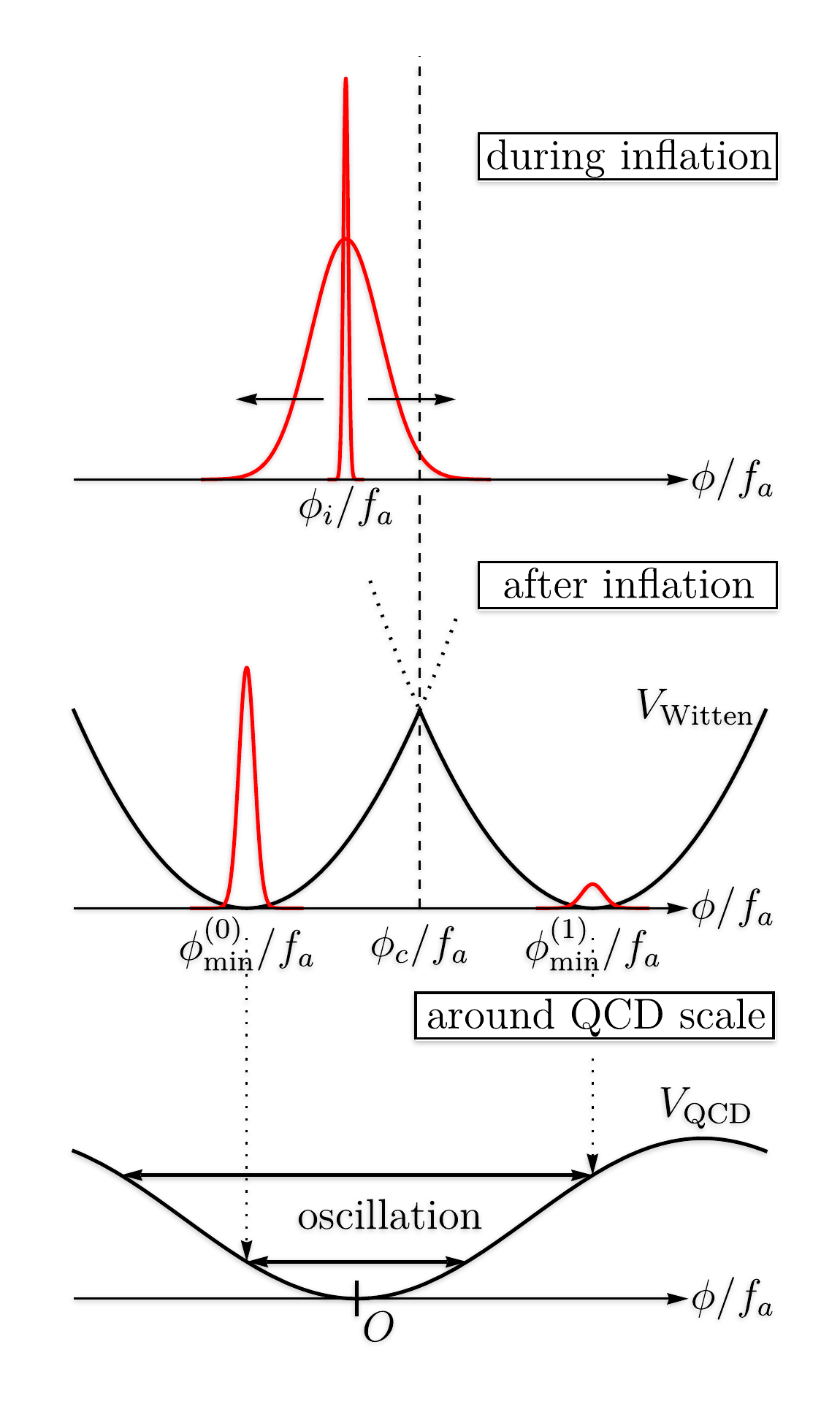}
    \caption{%
        Overview of the original scenario.
        The black curves denote the effective potential of the axion, and the red curves denote the probability distribution of the axion field.
        At first, a horizon patch corresponding to our observable universe has a certain field value, and it fluctuates during inflation.
        After inflation, the axion gets a periodic potential due to the Witten effect, and the universe separates into two regions with two different field values, which is determined by the field value at the end of inflation.
        The potential induced by the Witten effect disappears at the annihilation of monopoles and anti-monopoles before the QCD phase transition.
        Due to the QCD instanton effect, the axion takes on the potential around the QCD scale.
        The axion oscillation with two different amplitudes is realized in the two different regions.
    }
    \label{fig: explanation}
\end{figure}

Depending on the field value determined by the monopole-induced potential, $\phi$ oscillates with two different amplitudes, as shown in Fig.~\ref{fig: explanation}.
Since the number density of axions is determined by the oscillation amplitude, two different field values provide different axion number densities.
If the regions with $\phi_\mathrm{min}^{(1)}$ are rare, the dominant regions with $\phi_\mathrm{min}^{(0)}$ provide a homogeneous axion density, which can explain dark matter, while the rare regions with $\phi_\mathrm{min}^{(1)}$ become axion bubbles with local overdensity.
Note that we focus on the two minima, $\phi_\mathrm{min}^{(0)}$ and $\phi_\mathrm{min}^{(1)}$, and ignore the others due to their rareness.
\subsection{The mass of PBHs and AMCs}
\label{subsec: Statu}
Following Ref.~\cite{Kitajima:2020kig}, we calculate the axion abundance to discuss the properties of formed PBHs and their byproduct, AMCs.
At the onset of the background axion oscillation at $T\simeq T_\mathrm{osc}$, which satisfies $m_a(T_\mathrm{osc})=3H(T_\mathrm{osc})$, the axion number density, $n_a$, is obtained by
\begin{align}
    n_a(T_\mathrm{osc})
    =
    \frac{1}{2}\kappa m_a(T_\mathrm{osc})\phi_\mathrm{osc}^2F(\phi_\mathrm{osc})
    \ ,
    \label{eq: axion number}
\end{align}
where $\kappa=1.5$ is the numerical fudge factor, $\phi_\mathrm{osc}$ is the axion field value at the onset of oscillation, and $F(\phi)$ is the numerical correction called the anharmonic factor given by~\cite{Visinelli:2009zm}
\begin{align}
    F(\phi)
    =
    \left[1-\log{\left(1-\frac{(\phi/f_a)^2}{\pi^2}\right)}\right]^{1.16}
    \ .
\end{align}
Since an oscillating scalar field behaves like non-relativistic matter, it is convenient to use the number density divided by the entropy density $s(T_\mathrm{osc})$, which becomes constant.
The entropy density is written as
\begin{align}
    s(T_\mathrm{osc})
    =
    \frac{4\rho_\mathrm{rad}(T_\mathrm{osc})}{3T_\mathrm{osc}}
    =
    \frac{4M_\mathrm{Pl}^2H(T_\mathrm{osc})^2}{T_\mathrm{osc}}
    \ ,
\end{align}
where $\rho_\mathrm{rad}$ is the energy density of radiation, $M_\mathrm{Pl} \simeq 2.4 \times 10^{18}$\,GeV is the reduced Planck mass, and $H$ is the Hubble parameter.
Thus, for $T<T_\mathrm{osc}$, the axion-to-entropy ratio is given by
\begin{align}
    \frac{n_a(T)}{s(T)}
    =
    \frac{n_a(T_\mathrm{osc})}{s(T_\mathrm{osc})}
    =
    \frac{9\kappa T_\mathrm{osc}}{8 m_a(T_\mathrm{osc})}\left(\frac{\phi_\mathrm{osc}}{M_\mathrm{Pl}}\right)^2F(\phi_\mathrm{osc})
    \ .
    \label{eq: number density}
\end{align}
When the homogeneous QCD axion composes all dark matter, the oscillation amplitude is determined from Eq.~(\ref{eq: number density}) as
\begin{equation}
   \frac{\phi_\mathrm{min}^{(0)}}{f_a}
   =
   4.25 \times 10^{-3}
   \left( \frac{g_{\rm osc}}{60} \right)^{0.209}
   \left( \frac{f_a}{10^{16}\,\rm{GeV}} \right)^{-0.582}
   \ ,
\end{equation}
where $g_{\rm osc}$ is the effective relativistic degrees of freedom at the onset of oscillations due to $V_\mathrm{QCD}$.
On the other hand, the axion bubbles collapse into PBHs if their overdensities are large enough at the horizon entry. 
Otherwise, they form AMCs.
PBHs are formed when the axion bubbles are dominated by massive axions at their horizon reentry, which leads to the upper bound on the temperature at horizon reentry $T_B$ for the PBH formation~\cite{Kitajima:2020kig},
\begin{align}
\begin{split}
    T_\mathrm{B}
    &=
   3.04\,\mathrm{MeV}
   \left( \frac{g_{\rm osc}}{60} \right)^{-0.418}
   \left( \frac{f_a}{10^{16}\,\mathrm{GeV}} \right)^{1.16}
   \\
    &\quad\times
    \left[
        1
        +0.0842\ln\left( 
            \frac{f_a}{10^{16}\,\mathrm{GeV}} 
        \right)
        -0.0302\ln\left( 
            \frac{g_\mathrm{osc}}{60} 
        \right)
   \right]^{1.16}
   \ .
\end{split}
    \label{eq: matter-radiation}
\end{align}
This threshold temperature $T_\mathrm{B}$ is translated to the lower bound on the PBH mass, $M_\mathrm{min}$, as
\begin{align}
\begin{split}
    M_\mathrm{min}
    &=
    1.68 \times 10^4M_{\odot}
    \left(\frac{g_f}{10}\right)^{-\frac{1}{2}}
    \left( \frac{g_{\rm osc}}{60} \right)^{0.836}
    \left( \frac{f_a}{10^{16}\,\mathrm{GeV}} \right)^{-2.33}
    \\
    &\quad\times
    \left[
        1
        +0.0842\ln\left( 
            \frac{f_a}{10^{16}\,\mathrm{GeV}} 
        \right)
        -0.0302\ln\left( 
            \frac{g_\mathrm{osc}}{60} 
        \right)
   \right]^{-2.33}
   \ ,
\end{split}
    \label{eq: lower}
\end{align}
where $g_f$ is the effective relativistic degrees of freedom at PBH formation, while the PBH mass is evaluated by the background horizon mass at formation \cite{Kopp:2010sh, Carr:2014pga},
\begin{align}
    M_\mathrm{PBH}
    \simeq
    1.68\times 10^4M_\odot\left(\frac{g_f}{10}\right)^{-1/6}\left(\frac{k}{3.5\times 10^4\,\mathrm{Mpc^{-1}}}\right)^{-2}
    \ .
    \label{eq: PBH mass}
\end{align}
Here, $k$ is the wavenumber corresponding to the scale of the axion bubble.
We can consider $N$ as a function of the mass of PBHs $N(M)$ using Eqs.~\eqref{eq: PBH mass} since the e-folding number is defined by $N(k)\equiv\log(k/k_0)$ with $k_0=2.24\times 10^{-4}\,\mathrm{Mpc}^{-1}$ being the scale of the observable universe.
The e-folding number corresponding to the PBH with a mass of $M_\mathrm{min}$ is given by $N_\mathrm{PBH} \simeq 19$.
Note also that the PBH mass scales as $M_\mathrm{PBH} \propto e^{-2N}$, which will be used later.
On the other hand, the AMC mass is evaluated by the total axion mass inside the horizon at the horizon crossing as
\begin{align}
\begin{split}
    M_\mathrm{AMC}(k) 
    &=2.04\times 10^{-2} M_{\odot}
    \left(\frac{g_\mathrm{osc}}{60}\right)^{-0.418}
    \left(\frac{f_a}{10^{16}\,\mathrm{GeV}}\right)^{1.16}
    \left(\frac{k_\mathrm{osc}}{k}\right)^3
    \\
    &\quad\times
    \left[
        1
        +0.0842\ln\left( 
            \frac{f_a}{10^{16}\,\mathrm{GeV}} 
        \right)
        -0.0302\ln\left( 
            \frac{g_\mathrm{osc}}{60} 
        \right)
   \right]^{1.16}
    \ ,
    \label{eq: AMC mass}
\end{split}
\end{align}
for $k<k_\mathrm{osc}\equiv a(T_\mathrm{osc})H(T_\mathrm{osc})$ with $a(T_\mathrm{osc})$ being the scale factor at the onset of the axion oscillation around the QCD scale.
\subsection{Modified model}
\label{subsec: modif}
\begin{figure}[ht]
    \centering
    \includegraphics[width=.6\textwidth]{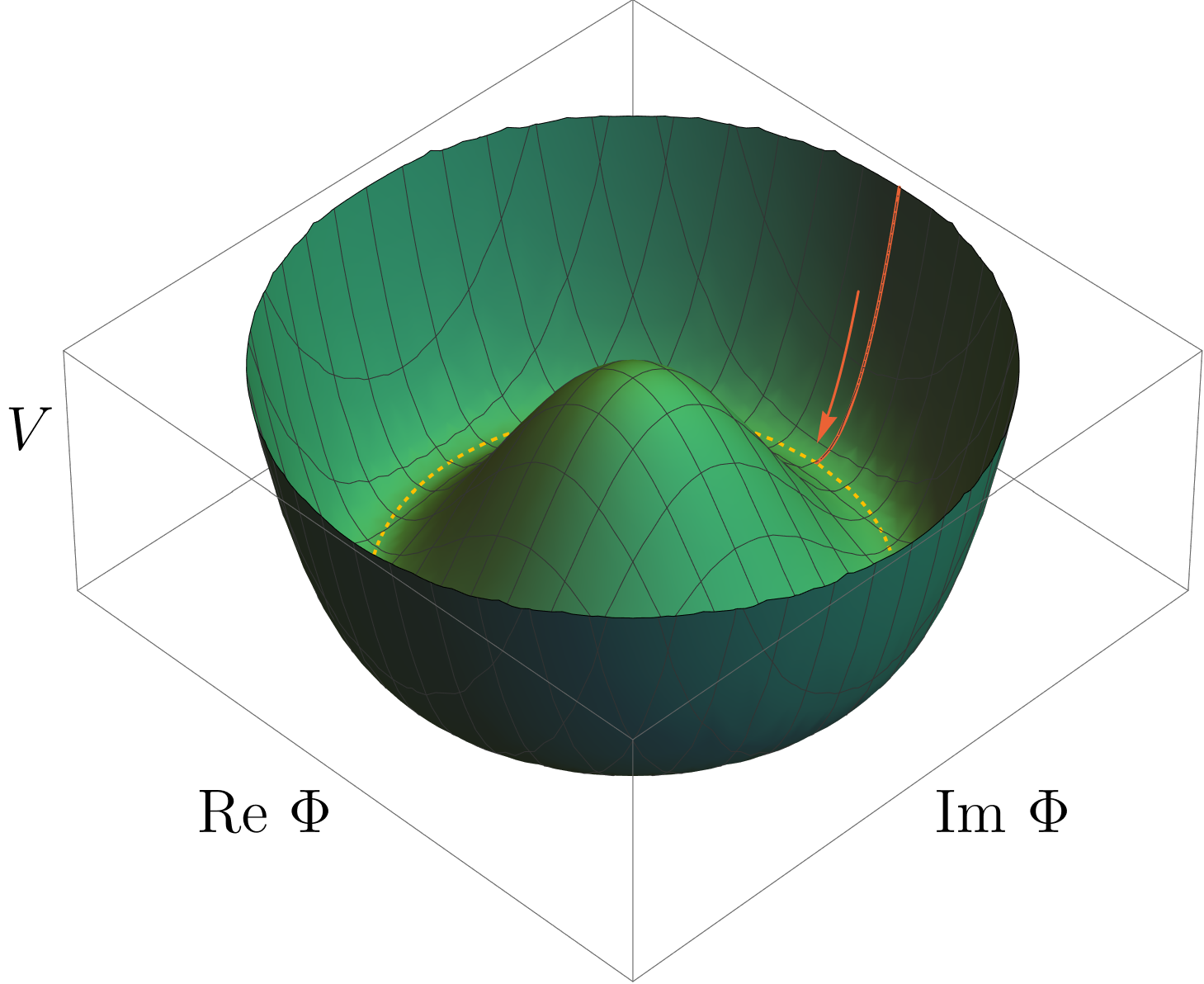}
    \caption{Schematic image of the dynamics of the PQ scalar during inflation. 
    In the early stage of inflation, the PQ scalar rolls down along the red curve.
    The PQ scalar finally reaches the potential minima in the radial direction, denoted by the dashed yellow circle.
    }
    \label{fig: rolling PQ}
\end{figure}

Next, we explain our modified setup and evaluate the abundance and clustering of PBHs and AMCs.
As we will see below, the modified model can reproduce the results of the original model by taking a certain limit.
Since we focus on the formation of PBHs accounting for the seeds of the SMBHs, we set the axion decay constant as $f_a=10^{16}\,\mathrm{GeV}$.
When the PQ field has a vacuum expectation value $f_a$, the probability density $P(N,\phi)$ that the axion field takes a value $\phi$ at the e-folding number $N$ during inflation evolves following the Fokker-Planck equation:
\begin{align}
    \frac{\partial P(N,\phi)}{\partial N} 
    =
    \frac{H_I^2}{8\pi^2}
    \frac{\partial^2 P(N,\phi)}{\partial \phi^2}
    \ ,
    \label{eq: Fokker-Planck}
\end{align}
where we assume that the Hubble parameter $H_I$ is constant during inflation for simplicity.
In this paper, we assume that the PQ scalar slowly rolls down from a large field value ($\gg f_a$) during inflation as shown in Fig.~\ref{fig: rolling PQ}. 
We denote the PQ scalar $\Phi$ in the polar coordinate as
\begin{align}
    \Phi
    =
    \frac{1}{\sqrt{2}}\varphi e^{i\theta},
    \label{eq: def of phi}
\end{align}
using the real variables, $\varphi$ and $\theta$, and assume that its potential at $|\Phi| \gtrsim f_a$ is dominated by the Hubble-induced mass term as
\begin{align}
    V(\varphi)
    \simeq
    \frac{1}{2}c_IH_I^2\varphi^2
    \qquad
    (\varphi \gtrsim f_a),
    \label{eq: potential}
\end{align}
where $c_I$ is an $\mathcal{O}(1)$ constant and is assumed to be positive here.
We then obtain the equation of motion for the radial part as
\begin{align}
    \ddot{\varphi}+3H_I\dot{\varphi}+c_IH_I^2\varphi
    =0,
    \label{eq: EoM}
\end{align}
where the dot($\cdot$) denotes the derivative with respect to the cosmic time $t$.
The solution is written as
\begin{align}
\begin{aligned}
    \varphi
    &=
    \varphi_\ast e^{-\lambda H_I(t-t_\ast)},
\\
    \lambda
    &=
    \frac{3}{2}\left(1\pm \sqrt{1-\frac{4}{9}c_I}\right),
\end{aligned}
    \label{eq: sol}
\end{align}
for $0<c_I<9/4$.
If $c_I > 9/4$, $\varphi$ experiences damped oscillations due to its large mass, which we do not consider in this work.
We focus on the minus sign in the solution of $\lambda$ because the solution with the plus sign damps faster.
The subscript $\ast$ means the value at $t=t_\ast$ defined as the time when $\varphi=\varphi_\ast\equiv f_a$. 
The radial field $\varphi$ should eventually settle at $\varphi = f_a$ due to the potential with its minimum at $f_a$.
A concrete realization of such a situation is given in Ref.~\cite{Kasuya:2009up}, and here we simply assume that $\varphi$ follows the solution~\eqref{eq: sol} for $t < t_*$ and stop at $\varphi = f_a$.
Since the scale factor is given by $a=a_\ast e^{H_I(t-t_\ast)}$ during inflation, the time evolution of the PQ scalar can be rewritten as
\begin{align}
    \varphi
    =
    \varphi_\ast \left(\frac{a}{a_\ast}\right)^{-\lambda}
    =
    f_a \left(\frac{a}{a_\ast}\right)^{-\lambda}
    \ .
    \label{eq: sol2}
\end{align}
Then, the e-folding number at $t=t_\ast$ satisfies
\begin{align}
    N_\ast
    =
    \ln{\frac{a_\ast}{a_i}}
    =
    \frac{1}{\lambda}\ln{\frac{\varphi_i}{\varphi_\ast}}
    =
     \frac{1}{\lambda}\ln{\frac{\varphi_i}{f_a}},
    \label{eq: Nstar}
\end{align}
where the subscript $i$ denotes the quantities at $N = 0$, i.e., the horizon exit of the current horizon scale during inflation.
In this paper, we assume $\varphi_i = M_\mathrm{Pl}$.
Note that the original scenario can be understood as the limit of $\lambda \to \infty$ in the modified scenario.%
\footnote{$\lambda \to 0$ also approaches to the original model with $f_a = \varphi_i$.}
For the typical value $\lambda=0.5$, we obtain $N_\ast\simeq 11$, which is smaller than $N_\mathrm{PBH}\simeq 19$ corresponding to the PBH with a mass of $M_\mathrm{min}$.
Thus, in the following, we consider only the case of
\begin{align}
    N_\mathrm{PBH}>N_\ast
    \ ,
    \label{eq: condition}
\end{align}
i.e., the radial field settles down to the minimum before the horizon exit of the PBH formation scale.
If $N_{\rm PBH} < N_*$, on the other hand, the PBH abundance is suppressed for the fixed initial field value. 
The PBH abundance can be adjusted by varying $H_I$ or $\varphi_i$, but our model modifications do not provide much benefit from suppression of isocurvature perturbations of AMC on large scales. 
Now, we are interested in the evolution of the phase component of the PQ scalar.
Since the fluctuations of the radial component, $\delta\varphi$, are damped due to the effective mass of $\mathcal{O}(H_I) $ during inflation, we can neglect them and consider that $\varphi$ homogeneously evolves following Eq.~\eqref{eq: sol2}.
The amplitude of the fluctuations of the phase component is determined by $H_I$.
Then, the fluctuations of $\theta$ are inversely proportional to $\varphi$.
For convenience, we define $\phi\equiv\theta f_a$, which is identified with the axion field when $\varphi = f_a$.
The Fokker-Planck equation for $\phi$ becomes 
\begin{align}
    \frac{\partial P(N,\phi)}{\partial N} 
    &=
    \frac{H_I^2}{8\pi^2}
    \left(\frac{f_a}{\varphi}\right)^2
    \frac{\partial^2 P(N,\phi)}{\partial \phi^2}
    \nonumber \\
    &=
    \frac{H_I^2}{8\pi^2}
    e^{2\lambda(N-N_\ast)}
    \frac{\partial^2 P(N,\phi)}{\partial \phi^2}
    \ ,
    \label{eq: rolling Fokker-Planck}
\end{align}
for $0 \le N \le N_*$.
Here, we define the following variable:
\begin{align}
    \tilde{N}(N)
    & \equiv
    \begin{cases}
        \frac{e^{2\lambda(N-N_\ast)}-1}{2\lambda}+N_\ast & \text{for $0 \leq N \leq N_*$\ ,}\\
        N & \text{for $N_* \leq N$\ .}
     \end{cases}
\end{align}
Then, the Fokker-Planck equation is written as 
\begin{align}
    \label{eq: Fokker-Planck_rev} 
    \frac{\partial \tilde{P}(\tilde{N},\phi)}{\partial \tilde{N}} 
    =
    \frac{H_I^2}{8\pi^2}
    \frac{\partial^2 \tilde{P}(\tilde{N},\phi)}{\partial \phi^2}
    \ ,
\end{align}
where $\tilde{P}(\tilde{N}, \phi)$ is defined by $\tilde{P}(\tilde{N}(N), \phi) = P(N, \phi)$.
Note that the above equation is valid for $N > N_*$ as well as $0 \le N \le N_*$.
By setting the initial condition $P(N=0,\phi) = \tilde{P}(\tilde{N}=\tilde{N}(0),\phi)=\delta(\phi-\phi_i)$, we obtain the solution of Eq.~\eqref{eq: Fokker-Planck_rev} as
\begin{align}
    \tilde{P}(\tilde{N},\phi ; \phi_i)
    =
    \frac{1}{\sqrt{2\pi}\sigma(\tilde{N})}
    e^{-\frac{(\phi-\phi_i)^2}{2\sigma(\tilde{N})^2}}
    \ , \quad
    \sigma(\tilde{N})
    \equiv 
    \frac{H_I}{2\pi}\sqrt{\tilde{N}-\tilde{N}(0)}
    \ .
    \label{eq: solution}
\end{align}
As discussed in Sec.~\ref{subsec: Revie}, if $\phi$ exceeds a threshold value, $\phi_c$, at the end of inflation, $\phi$ rolls down to $\phi_\mathrm{min}^{(1)}$ and form the axion bubbles after inflation.
Due to the periodicity, the threshold value is given by $\phi_c=(\phi_\mathrm{min}^{(0)}+\phi_\mathrm{min}^{(1)})/2$.
By integrating the probability distribution for $\phi > \phi_c$, we obtain the total volume fraction of the axion bubbles that eventually form PBHs or AMCs.
Then, the size distribution of the axion bubbles is given by
\begin{align}
    \beta_{N1}(N; \phi_i)
    =
    \frac{\partial}{\partial N}
    \int_{\phi_c}^\infty \mathrm{d}\phi \, \tilde{P}(\tilde{N}(N),\phi;\phi_i)
    =
    \frac{\phi_c-\phi_i}{2(N-\tilde{N}(0))}\tilde{P}(\tilde{N}(N),\phi_c;\phi_i)
    \ 
    \label{eq: formation prob}
\end{align}
for $N>N_*$,
where $N$ is 
related to the scale of 
the axion bubbles that form PBHs or AMCs.
The PBH energy density $\rho_\mathrm{PBH}$ is calculated from the background energy density at the formation and the volume fraction of the axion bubbles since the PBH mass is evaluated by the background horizon mass as mentioned before.
The PBH fraction in the energy density of dark matter, $f_\mathrm{PBH}$, is
\begin{align}
   \frac{{\rm d}f_{\rm{PBH}}}{{\rm d}\ln{M_\mathrm{PBH}}} 
   &\equiv
       \frac{1}{\rho_{\rm{DM},0}}
       \frac{{\rm d}\rho_{\rm{PBH}}}{{\rm d}\ln{M_\mathrm{PBH}}}
   \nonumber\\
   &
   =\frac{\rho_{\rm{rad}}(T_f) }{2\rho_\mathrm{DM}(T_f)} 
       \beta_{N1} (N(M_\mathrm{PBH}); \phi_i)
       \,\Theta(M_\mathrm{PBH}-M_\mathrm{min})
       \ .
   \label{eq: PBH fraction}
\end{align}
where $\rho_\mathrm{DM}$ denotes the energy density of dark matter, $T_f$ denotes the temperature at the PBH formation, and $\Theta$ is the Heaviside step function.
On the other hand, the AMC energy density, $\rho_\mathrm{AMC}$, is evaluated from the axion number density and the volume fraction of the bubbles.
The AMC fraction of the energy density of dark matter, $f_\mathrm{AMC}$, is
\begin{equation}
   \frac{{\rm d}f_{\rm{AMC}}}{{\rm d}\ln k} 
   \equiv
   \frac{1}{\rho_{\mathrm{DM},0}}\frac{{\rm d}\rho_{\mathrm{AMC}}(k)}{{\rm d}\ln{k}}
   =\frac{\rho_{\mathrm{in}}(k)}{\rho_{\mathrm{DM}}(T_\mathrm{osc})}
   \beta_{N1} (N(k);\phi_i)
   =
   \frac{n_a/s|_{\phi_\mathrm{osc}=\pi f_a-\phi^{(0)}_{\mathrm{min}}}}{n_a/s|_{\phi_\mathrm{osc}=\phi^{(0)}_{\mathrm{min}}}} 
   \beta_{N1} (N(k);\phi_i)
   \ ,
\end{equation}
for $k<k_\mathrm{osc}$, where $\rho_{\mathrm{in}}$ is the axion energy density inside the axion bubbles .
For $k>k_\mathrm{osc}$, since the bubbles reenter the horizon before the onset of the background axion oscillation, $\phi_\mathrm{osc}$ is damped compared to $\pi f_a - \phi_\mathrm{min}{(0)}$ inversely proportional to the scale factor.
By taking this effect into consideration, the AMC fraction is obtained as~\cite{Kitajima:2020kig, Kasai:2023ofh}
\begin{align}
    \frac{{\rm d}f_{\mathrm{AMC}}}{{\rm d}\ln k} 
    &=
    \frac{\tilde{n}_a/s|_{\phi_\mathrm{osc} = \phi_{\mathrm{dec}}}}{n_a/s|_{\phi_\mathrm{osc} =\phi^{(0)}_{\mathrm{min}}
    }} 
    \beta_{N,1} (N(k);\phi_i)
    \ ,
    \\
    \phi_{\rm dec}
    &\equiv(\pi f_a-\phi^{(0)}_{\mathrm{min}})
        \left(\frac{m_a (T_{\mathrm{osc}}) }{3H_k}\right)^{\frac{1}{2}},
   \label{eq: bubble fraction2}
\end{align}
where $H_k$ is the Hubble parameter at the horizon entry of the axion bubbles whose size is characterized by the wavenumber $k$.
Here, we define $\tilde{n}_a$ as the number density of relativistic axions, which is obtained by replacing $m_a(T_\mathrm{osc})$ with $\sqrt{m_a^2 (T_{\mathrm{osc}}) + k^2/a^2(T_{\rm osc})}$ in the definition of $n_a$, Eq.~\eqref{eq: axion number}.
Next, we consider the two-point correlation of the bubble formation following Ref.~\cite{Kawasaki:2021zir}.
The probability distribution of PBH formation at two points, $x$ and $y$, separated by a comoving distance $L$ is given by
\begin{align}
    \beta_{N2}(N_L,N_x,N_y)
    &=
    \frac{\partial}{\partial N_x}
    \frac{\partial}{\partial N_y}
    \int_{-\infty}^\infty \mathrm{d}\phi_L \, \tilde{P}(\tilde{N}(N_L),\phi_L;\phi_i)
    \nonumber \\
    &\int_{\phi_c}^\infty \mathrm{d}\phi_x \, \tilde{P}(\tilde{N}(N_x)-\tilde{N}(N_L),\phi_x;\phi_L)
    \int_{\phi_c}^\infty \mathrm{d}\phi_y \, \tilde{P}(\tilde{N}(N_y)-\tilde{N}(N_L),\phi_y;\phi_L)
    \ ,
    \label{eq: 2-pt prob 1}
\end{align}
where $N_{x,y}$ is the e-folding number corresponding to the PBHs formed at $x, y$, and $N_L$ is the e-folding number when the comoving distance $L$ exits the horizon.
As we will see below, the distributions of PBHs and AMCs have peaks at certain scales.
Thus, we approximate that the PBHs and AMCs have monochromatic mass distributions, and set $N_x=N_y=N$. 
We can then calculate a two-point correlation function for their formation with a separation $L$ by
\begin{align}
    \xi(L)
    =
    \frac{\beta_{N2}(N_L(L),N,N)}{\beta_{N1}(N)\beta_{N1}(N)}-1
    \ .
    \label{eq: correlation}
\end{align}
Using this correlation function, we can consider two observational constraints on this scenario.
One is on the isocurvature perturbations from the CMB observations, and the other is on the angular correlation function of the SMBHs from the quasar observations. 
\section{Observational constraints}
\label{sec: themo}
\subsection{PBH and AMC abundances and their correlations}
\label{subsec: PBHan}
\begin{figure}[t]
    \centering
    \subfigure[PBH abundance]{%
       \includegraphics[width=.5\textwidth ]{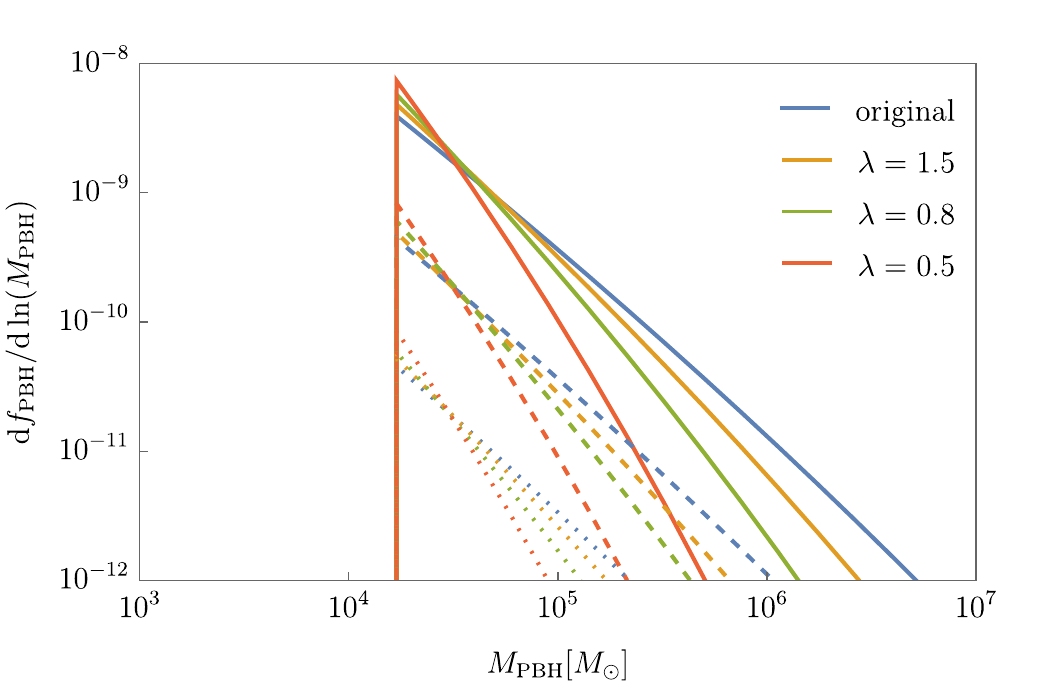}}%
    \subfigure[AMC abundance]{%
       \includegraphics[width=.5\textwidth ]{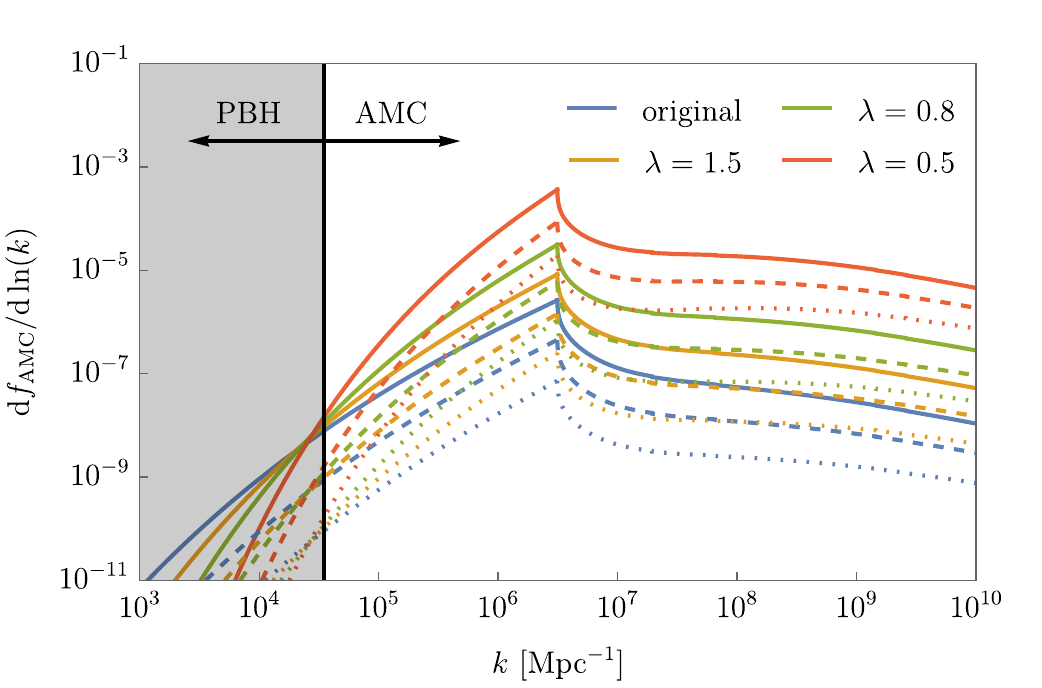}}%
    \caption{%
    Abundances of the PBHs and AMCs. 
    The solid, dashed, and dotted lines denote $f_\mathrm{PBH}=3\times 10^{-9}$, $3\times 10^{-10}$, and $3\times 10^{-11}$, respectively.
    To fix $f_\mathrm{PBH}$ among the different values of  $\lambda$, we adjust the initial axion field value, $\phi_i$.
    The vertical line in the right panel represents the minimum $k$ for the AMC formation.
    To the left of this line, PBHs are formed instead of AMCs.
    }
    \label{fig: abundance}
\end{figure}

\begin{figure}[t]
    \centering
    \subfigure[PBH correlation function]{%
       \includegraphics[width=.5\textwidth ]{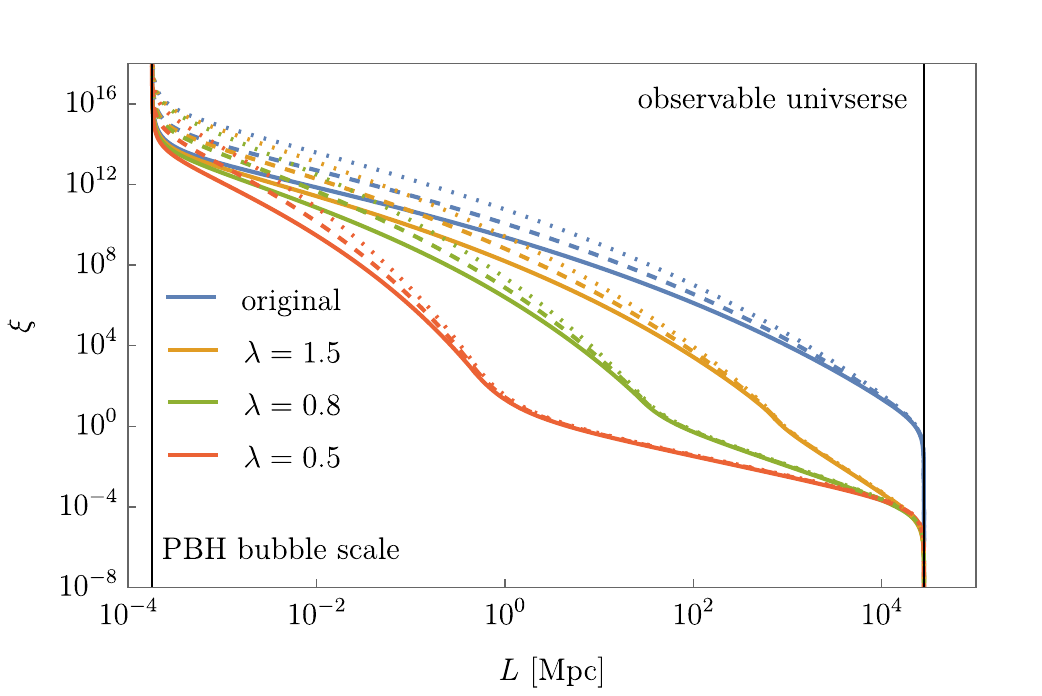}}%
    \subfigure[AMC correlation function]{%
       \includegraphics[width=.5\textwidth ]{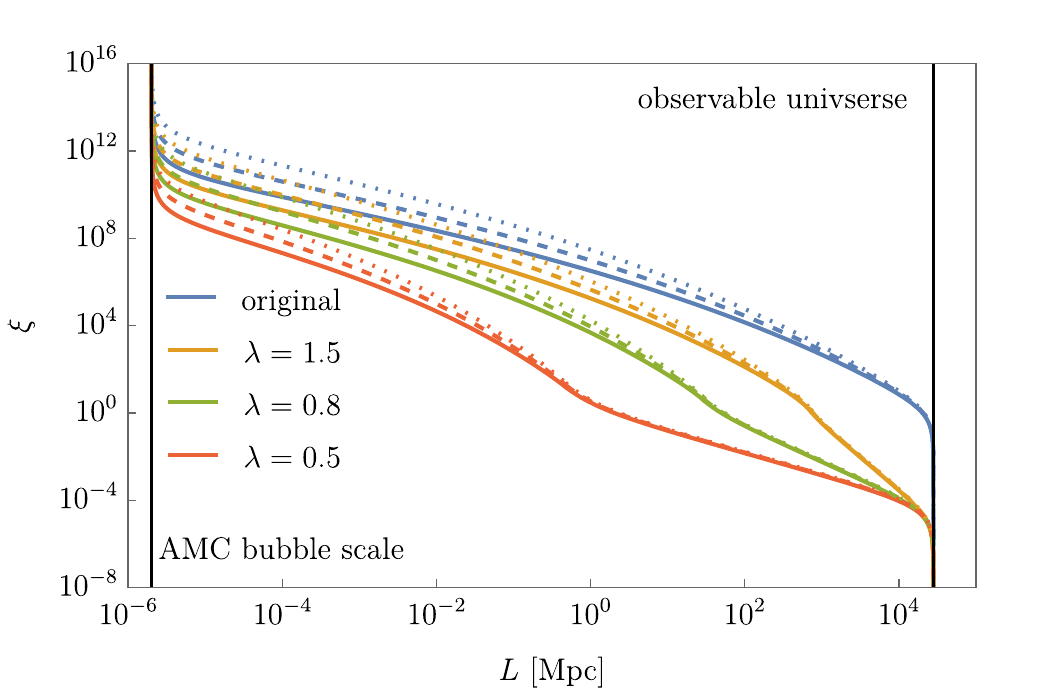}}%
    \caption{%
    Correlation functions of the PBHs and AMCs.
    The solid, dashed, and dotted lines denote $f_\mathrm{PBH}=3\times 10^{-9}$, $3\times 10^{-10}$, and $3\times 10^{-11}$, respectively.
    In both panels, the right vertical lines correspond to the size of the observable universe where $\xi \to 0$, while the left vertical lines correspond to the radius of the bubbles that form the PBHs or AMCs where $\xi \to \infty$.
    }
    \label{fig: correlation}
\end{figure}

We show the PBH and AMC abundances in Fig.~\ref{fig: abundance}. 
To explain the observed value $f_{\rm{SMBH}}\simeq 3\times 10^{-9}$~\cite{Willott:2010yu} in this model, we 
consider three different mass accretion onto the PBHs;
$M_{\rm{PBH}}$ increases due to the accretion by factors of $\mathcal{O}(1)$, $\mathcal{O}(10)$, and $\mathcal{O}(10^2)$.
They correspond to setting $f_\mathrm{PBH}=3\times 10^{-9}$, $3\times 10^{-10}$, and $3\times 10^{-11}$, respectively. 
Each value of $f_{\rm{PBH}}$ can be achieved by choosing an appropriate value of $\phi_c-\phi_i$ according to Eqs.~\eqref{eq: formation prob} and \eqref{eq: PBH fraction}.
The peak of the PBH mass distributions is at the mass cut-off, $M_\mathrm{PBH}=1.68\times 10^4M_\odot$ from Eq.~\eqref{eq: PBH mass}.
In Fig.~\ref{fig: abundance} (a), the mass distributions become steeper for smaller values of $\lambda$.
This is because, for smaller $\lambda$, the PQ scalar rolls down the potential longer, and the axion fluctuations are suppressed in a wider range of the scales.
In Fig.~\ref{fig: abundance} (b), we show the corresponding energy density distributions of the AMCs.
As in the PBH abundance, the AMC abundances also show steeper distributions for smaller $\lambda$.
The vertical line represents the minimum $k$ below which the PBHs are formed instead of the AMCs, with the area to its left being a gray-shaded region.
The peak corresponds to the scale reentering the horizon when the axion starts to oscillate, and then the peak mass is estimated as $M_\mathrm{AMC} \simeq 2.0\times 10^{-2}M_\odot$ from Eq.~\eqref{eq: AMC mass}.
To fix $f_\mathrm{PBH}$ in Fig.~\ref{fig: abundance}, we adjust $\phi_i$, which is the initial field value of axion, for each $\lambda$. 
In Fig.~\ref{fig: abundance}~(b), this results in the larger AMC fraction as $\lambda$ becomes smaller.
The PBH and AMC correlation functions are shown in Fig.~\ref{fig: correlation}.
Here, we approximated that the PBHs and AMCs have monochromatic mass distribution at the peak mass in Fig.~\ref{fig: abundance} and used the evaluation in Eq.~\eqref{eq: correlation}.
Due to the existence of the rolling phase of the PQ scalar, both the PBH and AMC correlation functions are suppressed, especially on large scales around the CMB scale.
The suppression of the correlation functions is more significant on broader scales for smaller $\lambda$ because the rolling phase lasts longer.
Since the size of the PBH ($\sim 2\pi k_\mathrm{PBH}^{-1}$) is much smaller than the observable universe scale, we redefine the correlation function by shifting the divergent point to the origin, which means we use $\xi(L-2\pi k_\mathrm{PBH}^{-1})$ as the correlation function in the following.

\subsection{Isocurvature perturbation}
\label{subsec: const}
\begin{figure}[t]
    \centering
       \includegraphics[width=.75\textwidth ]{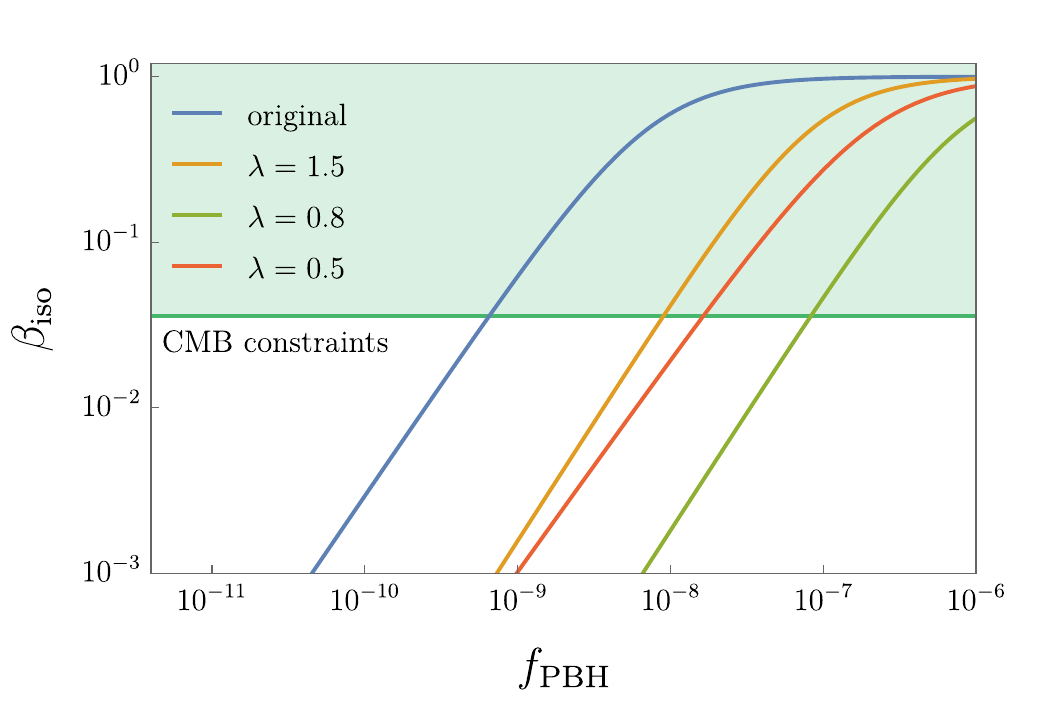}
    \caption{%
    Fraction of the isocurvature perturbations $\beta_\mathrm{iso}$ is shown as a function of the PBH fraction $f_\mathrm{PBH}$ for $\lambda =0.5, 0.8, 1.5$ and $\infty$ (original model).
    The green-shaded region is excluded from the CMB constraint on the isocurvature perturbations.
    }
    \label{fig:isocurvature}
\end{figure}

\begin{figure}[t]
    \centering
       \includegraphics[width=.75\textwidth ]{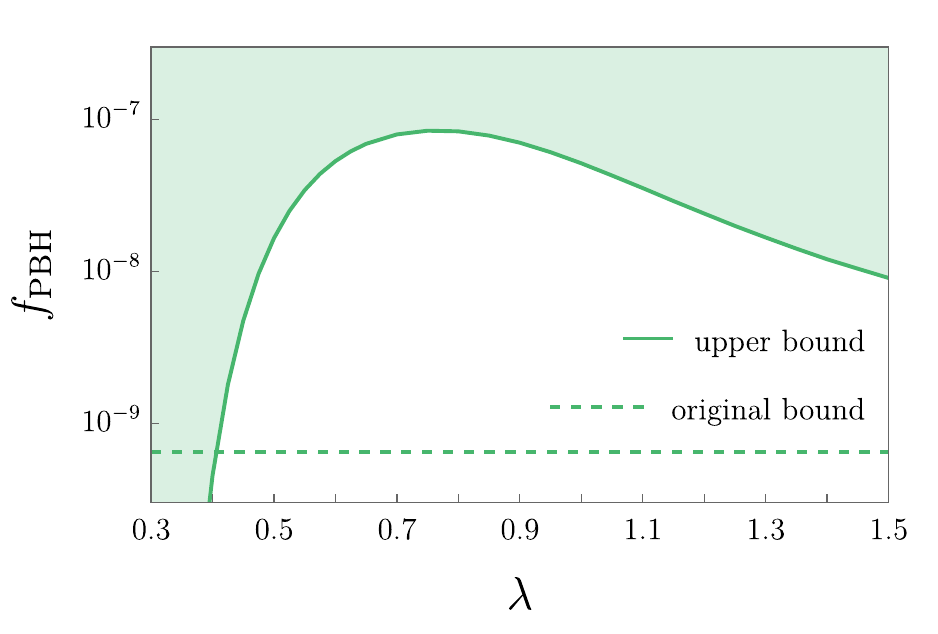}%
    \caption{%
    Upper bound on $f_\mathrm{PBH}$ from the CMB observation.}
    \label{fig:isocurvature_constraint}
\end{figure}

Using the correlation function, $\xi_\mathrm{PBH,AMC}$, we evaluate the dimensionless power spectrum of the density fluctuations of the PBHs or AMCs as
\begin{align}
    \mathcal{P}_\mathrm{PBH,AMC}(k)
    \equiv
    \left( \frac{k}{2\pi} \right)^3 
    \int {\rm d}^3x\,\xi_\mathrm{PBH,AMC}(x)e^{-i
    \bm{k}\cdot \bm{x}}
    \label{eq: P_xi integration}
    =
    \frac{k^3}{2\pi^2} \int {\rm d}r\,r^2\xi_\mathrm{PBH,AMC}(r)\frac{\sin{kr}}{kr}
    \ .
\end{align}
The PBHs contribute to the isocurvature perturbations through $f_\mathrm{PBH}^2\mathcal{P}_\mathrm{PBH}(k)$, while AMCs contribute to that through $f_\mathrm{AMC}^2\mathcal{P}_\mathrm{AMC}(k)$.
In fact, the PBH contribution is subdominant because $f_\mathrm{PBH}$ is much smaller than $f_\mathrm{AMC}$.
The Planck results~\cite{Planck:2018jri} give the upper bound on the isocurvature perturbations on the CMB scale as
\begin{equation}
   \beta_{\rm{iso}}(k_\mathrm{CMB})
   \equiv
   \frac{ \mathcal{P}_{\rm{iso}}(k_\mathrm{CMB}) }{ \mathcal{P}_{\rm{iso}}(k_\mathrm{CMB})+\mathcal{P}_{\mathcal{R}}(k_\mathrm{CMB}) }
   <
   0.036 \ ,
   \label{eq: obs const}
\end{equation}
where $\mathcal{P}_\mathrm{iso}$ is the dimensionless power spectrum of the isocurvature perturbation, $k_\mathrm{CMB} = 0.002\,\mathrm{Mpc}^{-1}$, and $\mathcal{P}_{\mathcal{R}}(k_\mathrm{CMB}) = 2\times 10^{-9}$ is the dimensionless power spectrum of the curvature perturbations on the CMB scale.
Using Eq.~\eqref{eq: P_xi integration}, we evaluate $\beta_\mathrm{iso}$ as a function of the PBH fraction $f_\mathrm{PBH}$ for a given $\lambda$, which is shown in Fig.~\ref{fig:isocurvature}.
In the figure, the constraint from the CMB observation is shown as the green-shaded region.
We also show the upper bound on $f_\mathrm{PBH}$ in Fig.~\ref{fig:isocurvature_constraint}.
For $\lambda\gtrsim 0.75$, the deviation of $\beta_{\rm{iso}}$ from the original model becomes larger when $\lambda$ becomes smaller, as expected from Eq.~\eqref{eq: Nstar}.
On the other hand, for $\lambda\lesssim 0.75$, the constraint on $f_{\rm{PBH}}$ from $\beta_{\rm{iso}}$ becomes more severe since the abundance of AMCs for fixed $f_{\rm{PBH}}$ significantly increases.
Specifically, for the scenario without mass accretion where  $f_\mathrm{PBH} \simeq 3 \times 10^{-9}$, $\lambda \gtrsim 0.45$ is required.

\subsection{Angular correlation}
\label{subsec: const2}
\begin{figure}[ht]
    \centering
    \includegraphics[width=.75\textwidth ]{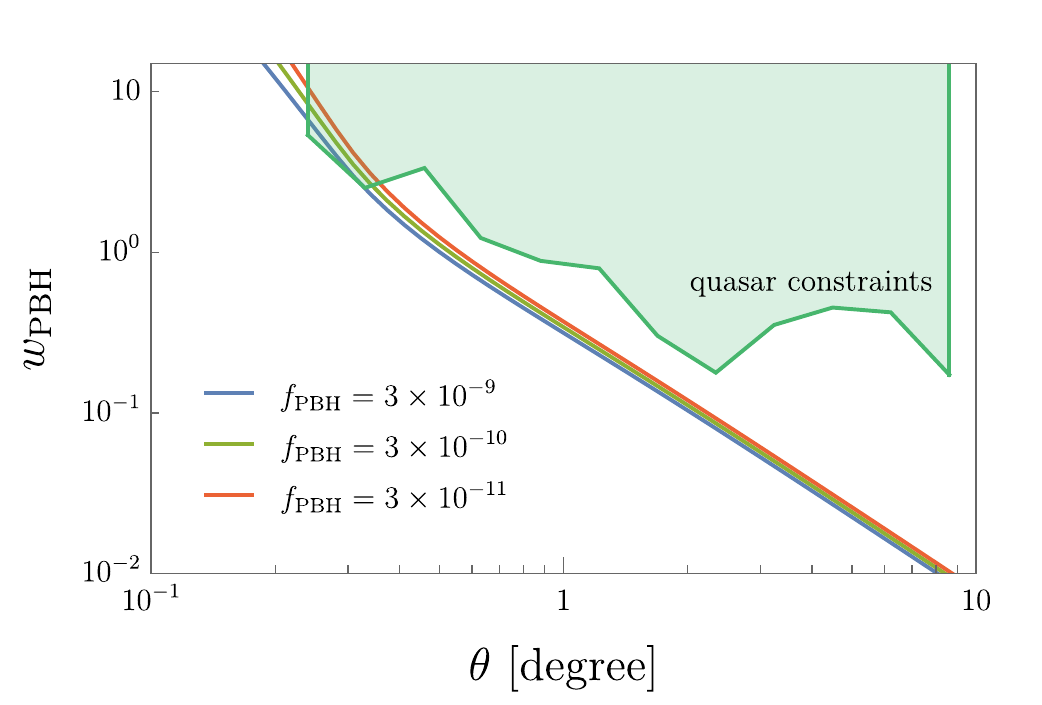}
    \caption{%
    Angular correlation function with fixed $\lambda=0.85$ and fixed $f_\mathrm{PBH}$.
    The green-shaded region denotes the $2\sigma$ constraints reported in Ref.~\cite{Shinohara:2023wjd}.
    Although there are variations in constraints, we can read more severe constraints at smaller angular separation.
    }
    \label{fig: angsep}
\end{figure}

\begin{figure}[ht]
    \centering
    \includegraphics[width=.75\textwidth ]{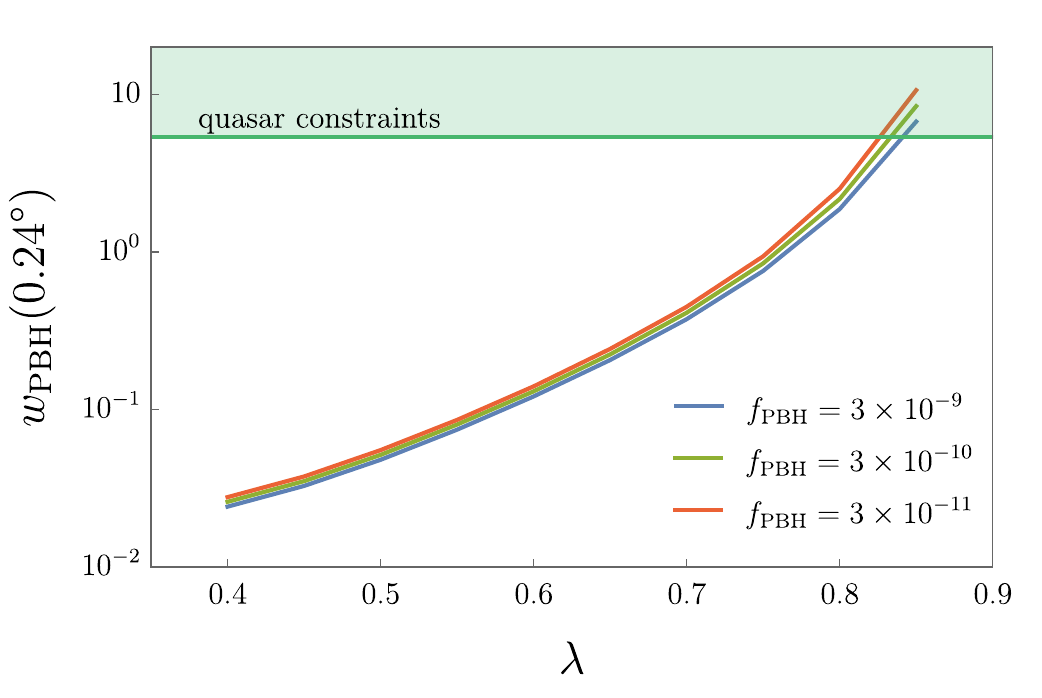}
    \caption{%
        Angular correlation of PBHs for $f_\mathrm{PBH} = 3 \times 10^{-9}$, $3 \times 10^{-10}$, and $3 \times 10^{-11}$.
        The green-shaded region is constrained by the observations of quasars.
        }
    \label{fig: angular}
\end{figure}

Here, we evaluate the angular correlation function of the PBHs following Ref.~\cite{Shinohara:2021psq}.
First, we set the comoving polar coordinate $(R,\theta,\phi)$ and the comoving number density of the PBHs, $n_\mathrm{PBH}(R,\theta,\phi)$, with the observer at the origin, $R = 0$.
Integrating along the line of sight, we obtain a two-dimensional number density as
\begin{align}
    N_\mathrm{PBH}(\theta,\phi)
    \equiv 
    \int_0^\infty \mathrm{d}R \,
    R^2 W(R) n_\mathrm{PBH}(R,\theta,\phi)
    \ ,
\end{align}
where $W(R)$ is the window function that limits the redshift range we observe.
Using the averaged number density, $\bar{n}_\mathrm{PBH}$, we also define the averaged two-dimensional number density by
\begin{align}
    \bar{N}_\mathrm{PBH}
    \equiv 
    \bar{n}_\mathrm{PBH}
    \int_0^\infty \mathrm{d}R \, R^2 W(R) 
    \ .
\end{align}
The number density contrast is defined by
\begin{align}
    \delta_\mathrm{PBH}(R,\theta,\phi)
    \equiv 
    \frac{n_\mathrm{PBH}(R,\theta,\phi)-\bar{n}_\mathrm{PBH}}{\bar{n}_\mathrm{PBH}}
    \ ,
\end{align}
which is equal to the energy density contrast for monochromatic PBHs.
Similarly, the two-dimensional number density contrast is defined by 
\begin{align}
    \Delta_\mathrm{PBH}(\theta,\phi)
    \equiv 
    \frac{N_\mathrm{PBH}(\theta,\phi)-\bar{N}_\mathrm{PBH}}{\bar{N}_\mathrm{PBH}}
    =
    \int_0^\infty \mathrm{d}R \,
    g(R) W(R) \delta_\mathrm{PBH}(R,\theta,\phi)
    \ ,
\end{align}
where
\begin{align}
    g(R)
    =
    \frac{R^2}{\int_0^\infty \mathrm{d}R^\prime \, {R^\prime}^2W(R^\prime)}
    \ .
\end{align}
The angular correlation function of the PBHs is defined by
\begin{align}
    w_\mathrm{PBH}(\theta)
    &\equiv 
    \langle \Delta_\mathrm{PBH}(\theta_1,\phi_1)\Delta_\mathrm{PBH}(\theta_2,\phi_2)\rangle
    \nonumber \\
    &=
    \int_0^\infty \mathrm{d}R_1  \int_0^\infty \mathrm{d}R_2 \,
    g(R_1) g(R_2) W(R_1) W(R_2)
    \langle 
        \delta_\mathrm{PBH}(R_1,\theta_1,\phi_1) 
        \delta_\mathrm{PBH}(R_2,\theta_2,\phi_2)
    \rangle
    \nonumber \\
    &=
    \int_0^\infty \mathrm{d}R_1 \int_0^\infty \mathrm{d}R_2
    g(R_1) g(R_2) W(R_1) W(R_2) \xi_\mathrm{PBH}(r)
    \ ,
\end{align}
where $\theta$ denotes the angular separation of the two directions given by $(\theta_1,\phi_1)$ and $(\theta_2,\phi_2)$, and $r \equiv \sqrt{R_1^2+R_2^2-2R_1R_2\cos{\theta}}$ is the distance of the two points, $(R_1, \theta_1,\phi_1)$ and $(R_2, \theta_2,\phi_2)$.
Using the top-hat form window function,
\begin{align}
    W(R) = \Theta(R - R_\mathrm{low}) \Theta(R_\mathrm{high} - R)
    \ ,
\end{align}
we obtain
\begin{align}
    w_\mathrm{PBH}(\theta)
    \equiv &
    \int_{R_\mathrm{low}}^{R_\mathrm{high}} \mathrm{d}R_1 
    \int_{R_\mathrm{low}}^{R_\mathrm{high}} \mathrm{d}R_2 
    \frac{3R_1^2}{R_\mathrm{high}^3-R_\mathrm{low}^3}
    \frac{3R_2^2}{R_\mathrm{high}^3-R_\mathrm{low}^3} \nonumber \\
    & \times \xi_\mathrm{PBH}\left( \sqrt{R_1^2+R_2^2-2R_1R_2\cos{\theta}} \right)
    \ ,
    \label{eq: angular correlation}
\end{align}
where $R_\mathrm{high}$ and $R_\mathrm{low}$ denote the distances to the maximum and minimum redshifts of the observations, $z_\mathrm{high}$ and $z_\mathrm{low}$, respectively.
Here we adopt $z_\mathrm{high} = 6.49$ and $z_\mathrm{low} = 5.88$ and consider a flat $\Lambda$CDM model following Ref.~\cite{Shinohara:2023wjd}.
According to Ref.~\cite{Shinohara:2023wjd}, the analysis of the quasars at $z\sim 6$ provides the upper bound on the two-point angular correlation function.
Since the angular correlation of PBHs at an angular separation of $\theta = 0.24^\circ$ gives the most stringent constraint, as shown in Fig.~\ref{fig: angsep}, we show the angular correlation function of the PBHs for $\theta = 0.24^\circ$,
and the 2$\sigma$ constraint, $w(0.24^\circ) < 5.37$, in Fig.~\ref{fig: angular}. 
The angular correlation becomes smaller for smaller $\lambda$.
This is because when the PQ scalar rolls down for a long time, it results in strong suppression of the PBH correlation function.
In the original model without the rolling of the PQ scalar, the angular correlation function is $w(0.24^\circ)=10^{6\,\text{--}\,7}$, which is much larger than the observational upper bound.
In Fig.~\ref{fig: angular}, one can see that our modification in this paper suppresses the angular correlation drastically.

\section{Discussion}
\label{sec: concl}
In this paper, we have studied a modification of the PBH formation model from axion bubbles.
We assume that the PQ scalar rolls down in the radial direction during inflation, which suppresses the isocurvature perturbations on large scales and weakens the CMB constraint from them.
While the lower bound of $\lambda$ is obtained by CMB, angular correlation functions of SMBHs provide the upper bound of $\lambda$.
As a result, for $f_\mathrm{PBH}=3\times10^{-9}$, $0.45\lesssim\lambda\lesssim0.8$ can avoid the observational constraints, which corresponds to $1.15\lesssim c_I\lesssim 1.76$.
This result shows that the abundance of SMBHs can be explained by PBHs even without accretion. 
On the other hand, in terms of each SMBH mass, this case assumes frequent PBH mergers to achieve the SMBH mass.
As one can see in Fig.~\ref{fig: correlation}, since our model has large PBH correlations on small scales, frequent PBH mergers are naively expected.
Then the assumption looks appropriate for our model, but we still need a detailed calculation of the PBH merger rate to test this model.
We have assumed that the initial value of the PQ scalar is as large as the Planck scale in this paper.
For a smaller initial value, the suppression of the axion fluctuations is weaker, and hence the constraint becomes more stringent.
For example, if we take $\varphi_i=0.1M_\mathrm{Pl}$ and $f_\mathrm{PBH}=3\times 10^{-9}$, we obtain $0.3\lesssim\lambda\lesssim 1.5$ from the isocurvature constraint and $\lambda\lesssim 0.4$ from the quasar angular correlations.
Thus, we conclude that our model can avoid observational constraints even with a sub-Planckian $\varphi_i$, although more stringent constraints are put on such cases.
\begin{acknowledgments}
This work was supported by JSPS KAKENHI Grant Nos. 20H05851(M.K., N.K., and F.T.), 21K03567(M.K.), 23KJ0088 (K.M.), 20H01894 (F.T. and N.K.), 21H01078 (N.K.), and 21KK0050 (N.K.), JSPS Core-to-Core Program (grant number: JPJSCCA20200002) (F.T.), and JST SPRING (grant number: JPMJSP2108) (K.K.).
K.K. was supported by the Spring GX program.
This article is based upon work from COST Action COSMIC WISPers CA21106, supported by COST (European Cooperation in Science and Technology).
\end{acknowledgments}

\bibliographystyle{JHEP}
\bibliography{Ref}

\end{document}